\documentclass{iaus}
\usepackage{graphicx}

\title[Maser polarization and magnetic fields] 
{A review of maser polarization and magnetic fields}

\author[W.~Vlemmings]   
{W.~H.~T.~Vlemmings$^{1,2}$}

\affiliation{$^1$Argelander Institute for Astronomy, University of Bonn, Auf dem H{\"u}gel 71, 53121 Bonn, Germany\\[\affilskip]
$^2$Jodrell Bank Observatory, Univ. of Manchester, Macclesfield, Cheshire SK11 9DL, U.K. \break email: Wouter.Vlemmings@manchester.ac.uk
}

\pubyear{2007}
\volume{242}  
\pagerange{119--126}
\date{?? and in revised form ??}
\jname{Astrophysical masers and their environments}
\editors{J.M. Chapman \& W.A. Baan, eds.}
\begin{document}

\maketitle

\begin{abstract}

Through polarization observations masers are unique probes of the
magnetic field in a variety of different astronomical objects, with
the different maser species tracing different physical conditions. In
recent years maser polarization observations have provided insights in
the magnetic field strength and morphology in, among others, the
envelopes around evolved stars, Planetary Nebulae (PNe), massive
star forming regions and supernova remnants. More recently, maser
observations have even been used to determine the magnetic field in
megamaser galaxies. This review will present an overview of maser
polarization observations and magnetic field determinations of the
last several years and discuss the implications of the magnetic field
measurements for several important fields of study, such as aspherical
PNe creation and massive star formation.

\end{abstract}

\firstsection 
\section{Introduction}

Because of their compactness, their high brightness and the fact that
they occur in a wide variety of astrophysical environments, masers are
excellent astrophysical probes. As the masers are often highly
linearly and circularly polarized, polarization observations add
fundamental information on both the masing process (such as pumping
and level of maser saturation) and the physical conditions in the
masing gas. With a detailed theory of maser polarization propagation,
the observations can yield the strength of the magnetic field along
the maser line of sight and the two-dimensional (or even
three-dimensional) field structure. Consequently, they allow for a
determination of the dynamical importance of the magnetic field. And,
since masers are specifically suited for high angular resolution
observations with interferometry instruments, the polarization
observations can probe the magnetic field properties at unprecedented
small scales.

In the period since the previous IAU maser conference (Migenes \& Reid
2002), the number of maser polarization observations have increased
significantly. They have been used to determine the strength and
structure of magnetic field in the circumstellar envelopes (CSE) of
Asymptotic Giant Branch (AGB) stars, high-mass star forming regions
and supernova remnants (SNRs), while observations of the magnetic
field strength in megamasers have also become possible. This review
details specifically the results on magnetic field measurements
derived from maser polarization observations since 2001, focusing on
the maser transitions in the radio-wavelength regime. In
\S\ref{sect1}, the theoretical background of maser polarization and
analysis considerations are discussed briefly. Recent magnetic field
measurements and their consequences for related astrophysical problems
involving evolved stars and high-mass star forming regions are
presented in \S\ref{sect2} and \S\ref{sect3}
respectively. \S\ref{sect4} details observations of SNR and megamaser
magnetic fields and finally \S\ref{sect5} presents perspectives for
future instruments and observations.

\section{Background and considerations}
\label{sect1}

There exists extensive literature on the theory of maser polarization,
as polarization during maser amplification differs from the regular
thermal emission case due to the stimulated emission process and a
range of other properties of the masing process that can influence the
radiation polarization characteristics. The main theoretical problem
is framed by constructing the density matrix evolution and radiative
transfer equations for the maser emission including the Zeeman terms
(\cite[Goldreich et al. 1973]{Goldreich73}). The Zeeman effect occurs
when the degeneracy of magnetic substates is broken under the
influence of a magnetic field. The magnitude of the Zeeman effect is
significantly different for paramagnetic (e.g. OH) and
non-paramagnetic molecules (e.g. SiO, H$_2$O and methanol), due to the
ratio between the Bohr magneton ($\mu_B = e\hbar / 2m_ec$) and the
nuclear magneton ($\mu_N = e\hbar/ 2m_nc$). In these expressions $e$
is the electron charge, $\hbar$ is the Planck constant and $c$ the
speed of light. The ratio of the two ($\mu_B / \mu_N \approx 10^3$) is
determined by the ratio of the electron mass ($m_e$) and the nucleon
mass ($m_N$) and implies three orders of magnitudes larger Zeeman
splitting for a similar magnetic field strength in the case of
paramagnetic molecules compared to the non-paramagnetic ones.

A fundamental difference in the treatment of polarized maser emission
exists between cases when the magnetic transitions overlap in
frequency or when they are well separated. This can be defined by the
splitting ratio $r_Z = \Delta\nu_Z / \Delta\nu_D$, where $\Delta\nu_Z$
is the Zeeman splitting and $\Delta\nu_D$ the Doppler
line width. Typically, $r_Z \gtrsim 1$ for the paramagnetic molecule
OH, while $r_Z < 1$ for the other, non-paramagnetic maser species. In
the case of $r_Z > 1$ there are no theoretical ambiguities and the
Zeeman components are well separated and resolved. The magnetic
transitions $\Delta m_F = \pm 1$ give rise to the $\sigma^{\pm}$
components, circularly polarized perpendicular to the magnetic field
$B$. The transition $\Delta m_F=0$ gives rise to the $\pi$ component,
linearly polarized along $B$. For an arbitrary angle between $B$ and
the maser propagation direction $\theta$, the resultant components are
elliptically polarized for $\theta < \pi/2$, and linearly polarized for
$\theta = \pi/2$. The observed splitting of the Zeeman components
directly gives the magnetic field strength $B\cos\theta$.

In the case of $r_Z < 1$, the Zeeman components overlap. Theoretical
work has been done both analytically and numerically with different
implications for the derived magnetic field strengths
(e.g.\cite[Watson 1994]{Watson94}, \cite[Elitzur 1996]{Elitzur96}, and
references therein). A comparison of the different polarization
theories is given by \cite{Gray03}, who finds the numerical models can
be used to accurately describe the maser polarization. This is further
supported by H$_2$O maser polarization observations that are best
described by the numerical models by \cite{Nedoluha92}. When $r_Z < 1$
the relation between the linear polarization angle and the magnetic
field direction is complex. The magnetic field direction is either
parallel or perpendicular to the magnetic field, depending on
$\theta$. When $\theta<\theta_{\rm crit}\approx 55^\circ$, the
polarization vectors are parallel to the magnetic field and when
$\theta>\theta_{\rm crit}$ they are perpendicular. However, this
relation is only valid when the Zeeman frequency shift under the
influence of the magnetic field $g\Omega$ is much larger than the rate
of stimulated emission $R$. Otherwise, the relation between
polarization angle and magnetic field direction is dependent on the
maser intensity. However, this is unlikely to affect any except the
strongest of maser features. The predicted relation between $\theta$
and the polarization angle from H$_2$O maser theory is supported by
the observations of a $90^\circ$ polarization angle flip when
$\theta$ crosses the critical angle in combination with the predicted
decrease in linear polarization fraction (\cite[Vlemmings \& Diamond
  2006]{Vlemmings06d}).

The derived $B$ strengths when $r_Z < 1$ do not only depend on the
circular polarization fraction but also depend on the maser saturation
level. Especially for masers that are saturated, simple assumptions
with a fixed proportionality between circular polarization and magnetic
field strength can lead to $B$ being overestimated by up to a factor
of 4. On the other hand, velocity gradients along the maser path can
cause the true field strength to be underestimated by a factor of two
(\cite[Vlemmings 2006]{Vlemmings06}). Also the blending of maser
features in lower angular resolution observations typically causes
another factor of two underestimation of $B$ (\cite[Sarma et
  al. 2001]{Sarma01}).

In both cases ($r_Z < 1$ and $r_Z > 1$) there are several other
properties of the maser and its surrounding medium that need to be
taken into account when interpreting polarization
observations. Especially at the low frequencies, Faraday rotation can
make a direct connection between the polarization angle and the
$B$-field uncertain. External Faraday rotation can cause significant
vector rotation originating along the line of sight to the maser
source. For instance at 1.6~GHz, a typical interstellar electron
density and magnetic field strength can cause up to a full $\sim
180^\circ$ rotation towards the W3(OH) star forming
region. Additionally, internal Faraday rotation can alter the
polarization characteristics of individual maser features in a source
in defferent ways, possibly destroying any large scale structure in
the linear polarization measurements (e.g. \cite[Fish \& Reid
  2006]{Fish06}).

\section{Evolved stars and planetary nebulae}
\label{sect2}

Maser polarization observations are the predominant source of
information about the role of magnetic fields during the late stages
of stellar evolution. Most observations have focused on the masers in
the CSEs of AGB stars, as OH, H$_2$O and SiO masers are fairly common
in these sources. However, polarization observations of masers around
post-AGB stars and (Proto-)PNe are becoming more common as more such
sources with maser emission are found (see G{\'o}mez, 2007, these
proceedings).

\subsection{AGB stars}

\begin{figure}
 \centerline{ \scalebox{0.32}{\includegraphics{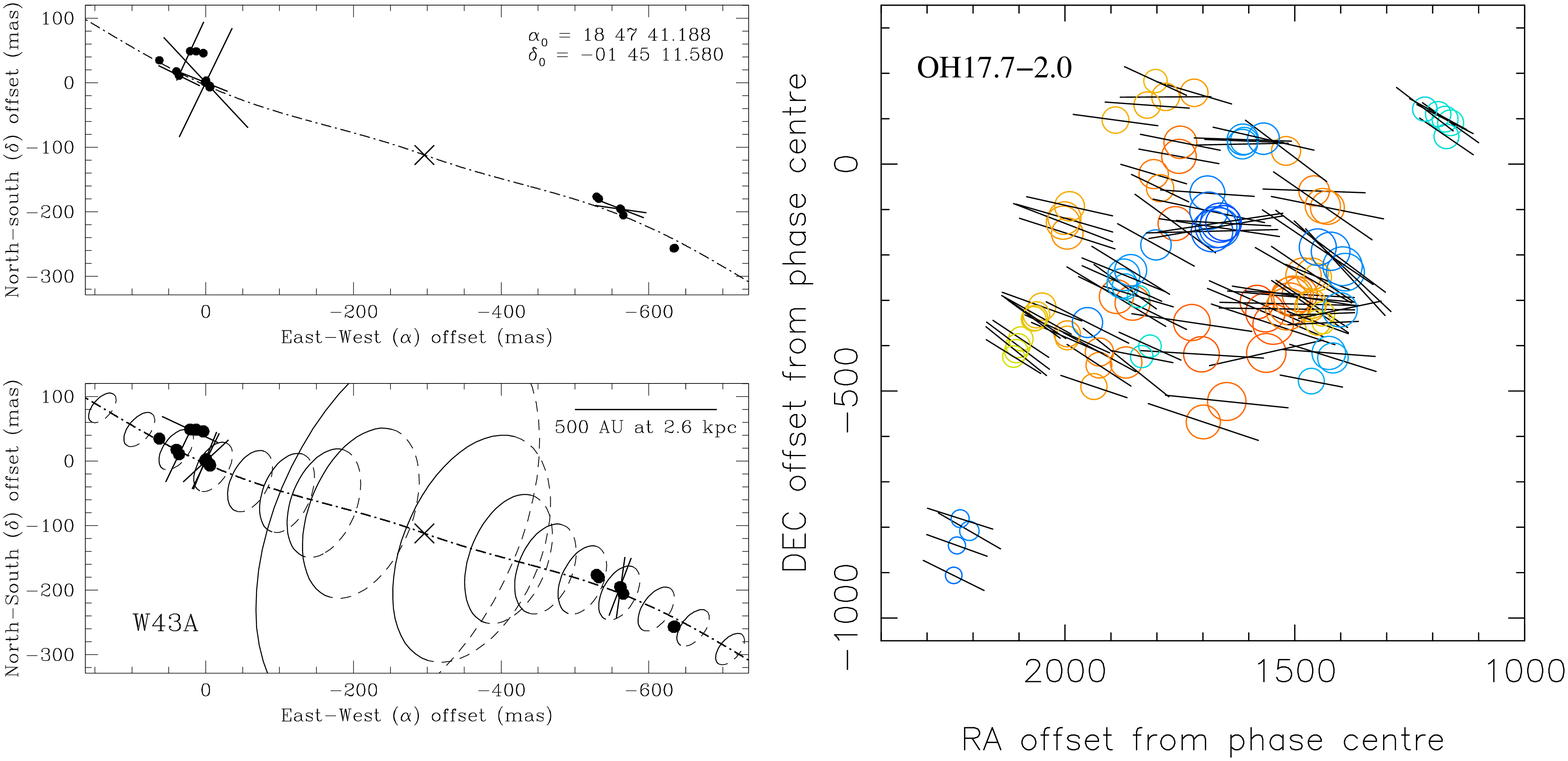}}}
  \caption{(left; top) The H$_2$O masers in the precessing jet
    (dashed-dotted line) of W43A (indicated by the cross). The maser
    features with the determined linear polarization vectors scaled
    linearly according to the fractional linear polarization. The
    polarization vectors lie predominantly along the jet with a median
    angle of $\chi = 63\pm12^\circ$ east of north. (left; bottom) The
    toroidal $B$-field of W43A. The vectors indicate the determined
    direction of $B$, perpendicular to the polarization vectors, at
    the location of the H$_2$O masers. The ellipses indicate the
    toroidal field along the jet, scaled with $B\propto r^{-1}$
    (\cite[Vlemmings et al. 2006a]{Vlemmings06a}). (right) The spatial
    distribution of the 1612 MHz linearly polarized maser components
    around the P-PNe OH17.7-2.0. Symbol size is proportional to Stokes
    I flux density; vector size is proportional to logarithmically
    scaled linearly polarized flux density. More blue-shifted
    components are represented by darker (blue) symbols and more
    red-shifted are lighter (red). The clear, overall regular magnetic
    field structure is consistent with a stretched dipole field
    (\cite[Bains et al. 2004]{Bains04})}
\label{Fig:fig1}
\end{figure}

\begin{figure}
 \centerline{ \scalebox{0.42}{\includegraphics{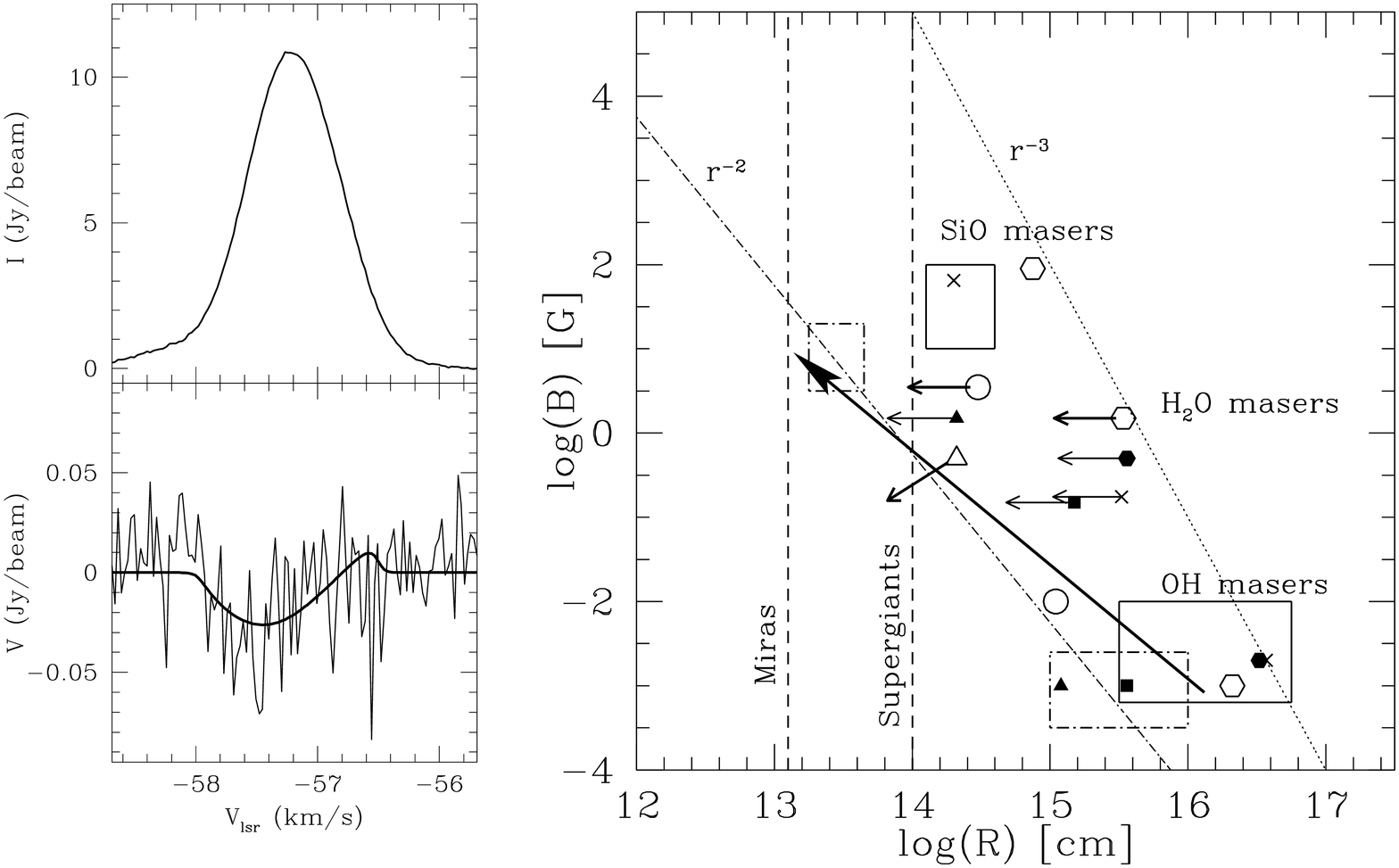}}}
  \caption{(left) The total intensity and circular polarization
    spectrum of the H$_2$O maser feature of W43A for which circular
    polarization was detected corresponding to a magnetic field
    strength of $85$~mG (\cite[Vlemmings et
      al. 2006a]{Vlemmings06a}). (right) The figure reproduced from
    \cite{Vlemmings05} of measured $B$-fields on the masers in the
    CSEs of evolved stars. The dashed-dotted boxes indicate the range
    of magnetic field strengths measured on the SiO and OH masers of
    Mira stars and the solid boxes those of Supergiant stars. The thin
    arrow indicate the H$_2$O maser $B$-fields measured in
    \cite{Vlemmings02} and \cite{Vlemmings05}, where the length of the arrows
    indicate the thickness of the H$_2$O maser shell with the symbols
    drawn on the outer edge. The symbols without arrows are the
    measurements on SiO and OH masers from the literature on the same
    sample of stars. The thick solid arrow indicates $B$ in the jet of
    W43A.}
\label{Fig:spec}
\end{figure}

While polarization observations of CSE 1.6~GHz OH masers are fairly
commonplace, recent years have seen an increase in 22~GHz H$_2$O and
43~GHz SiO maser observations of mainly Mira variables and
supergiants. As the different maser species typically occur in
different regions of the CSE, combining observations of all three
species allows us to form a more complete picture of the magnetic
field throughout the entire envelope. Close to the central star, SiO
maser linear polarization reveals an ordered $B$-field with a linear
polarization fraction ranging from $m_l \sim 15\%$ to $m_l \sim 65\%$
(e.g. \cite[Kemball \& Diamond 1997]{Kemball97}). Strikingly, radially
elongated jet-like SiO maser structures ,that have been observed
around for example $o$~Ceti and R~Aql, are apparently aligned with the
$B$-field (\cite[Cotton et al. 2006]{Cotton06}). A recent large single
dish survey of SiO maser polarization revealed an average field
strength of 3.5~G when assuming a regular Zeeman origin of the
polarization, indicating a dynamically important $B$-field
(\cite[Herpin et al. 2006]{Herpin06}). The observations find no
specific support for other (non-Zeeman) interpretations of the
polarization.

Further out in the envelope, also 22~GHz H$_2$O maser measurements reveal
significant $B$-fields, both around Miras and supergiants
(\cite[Vlemmings et al. 2005, and references therein]{Vlemmings05}). The
measured field strength is typically of the order of $\sim 100-300$~mG
but can be up to several Gauss. The strongest field strengths are
found around Mira variables, consistent with the H$_2$O masers
occurring closer to the star. As no linear polarization has been
detected thus far, describing the magnetic field shape is
difficult. However, for the Supergiant VX~Sgr, the complex maser
structure reveals an ordered field reversal across the maser region
consistent with a dipole $B$-field. Interestingly, the
orientation of the field determined from the H$_2$O maser polarization
is similar to the orientation of the outflow determined from other
H$_2$O maser observations as well as the orientation of a dipole field
determine from OH maser polarization (\cite[Vlemmings et
  al. 2005]{Vlemmings05}).

Finally, at typically even larger distances, 1.6~GHz OH maser polarization are
fairly common. Often strongly linearly polarized, the masers reveal an
ordered $B$-field at several thousands AU from the star. The
field strength at these distances is typically of the order of a few
mG (e.g. Deacon et al. 2007, in prep.; \cite[Etoka \& Diamond
  2004]{Etoka04}).

\subsection{Proto-Planetary nebulae}

Most of the polarization work of P-PNe is done on OH masers. Similar
to the magnetic field strengths around their progenitor stars, the
P-PNe fields are $\sim 1$~mG in the OH maser region. Single dish
surveys reveal linear and circular polarization in respectively $\sim
50\%$ and $\sim 75\%$ of the sources, dependent on the OH maser line
(polarization is more common in the 1612~MHz OH satellite line than in
the 1665 and 1667 main line masers). The polarization fraction is
typically less than $15\%$ (\cite[Szymczak \& G{\'e}rard 2004
]{Szymczak04}). The OH maser polarization has also been mapped using
MERLIN. As shown in Fig.~\ref{Fig:fig1}(right), the linear
polarization vectors reveal a highly ordered $B$-field
(\cite[Bains et al. 2003 \& 2004]{Bains03}).

A very small fraction of the Post-AGB/P-PNe maser stars show highly
collimated H$_2$O maser jets (see Imai 2007; these proceedings). These
so-called water-fountain sources are likely the progenitors of bipolar
PNe and there are indications that they evolve from fairly high-mass
AGB stars. The archetype of this class is W43A (\cite[Imai et
  al. 2002]{Imai02}) and polarization observations have recently
revealed that the maser jet is magnetically collimated
(Fig.\ref{Fig:fig1} and \ref{Fig:spec}; \cite[Vlemmings et
  al. 2006a]{Vlemmings06a}). This lends strong support to the theories
of magnetic shaping of PNe (e.g. \cite[Garc{\'{\i}}a-Segura et
  al. 2005]{Garcia05}). In addition to the observations of W43A,
recent Australia Telescope Compact Array (ATCA) observations of the
likely water-fountain source IRAS~15445-5449 (Vlemmings \& Chapman 2007, in
prep.) also indicate a magnetic H$_2$O maser jet.

\subsection{Planetary Nebulae}

There are only a handful of PNe known which show maser emission, and
even less of these have masers that are strong enough to provide
$B$-field measurements from polarization observations. One of the
sources that shows both OH and H$_2$O maser emission is the very young
PNe K3-35. In this source, the OH masers indicate a $B$-field of
a few mG at $~800$~AU from the central object (\cite[G{\'o}mez et
  al. 2006]{Gomez06}).

\subsection{Summary}

Fig.~\ref{Fig:spec}(right) gives a summary of the current magnetic
field measurements in CSEs of both Mira and supergiants. Although the
exact relation between the magnetic field strength and distance to the
central star remains uncertain, the field strengths are obvious strong
enough to dynamically influence the shaping of the outflow and help
shape asymmetric PNe. This is further supported by recent observations
of strong $B$-fields on the surface of the central stars in
several PNe (\cite[Jordan et al. 2005]{Jordan05}) as well as recent
PNe dust polarization observations (e.g. \cite[Sabin et
  al. 2007]{Sabin07}). The magnetic field could also the missing
component in the stellar mass-loss mechanism, as recent models
indicate the pulsation and radiation pressure alone might not be
enough (\cite[Woitke 2006]{Woitke06}). However, the question of the
origin of the magnetic field remains and will likely have to be sought
in the interaction with a heavy planet, a binary companion or a
circumstellar disk, although several models also claim a magnetic
dynamo can be maintained by a single star.

\section{High-mass star formation}
\label{sect3}

Star forming regions, and especially those forming high-mass stars,
often contain a wide variety of maser species tracing many different
density and temperature regimes. As was the case for evolved stars,
most information on the small scale magnetic fields comes from maser
polarization observations, dominated by OH maser measurements but also
with an increasing number of H$_2$O and methanol maser
observations. As SiO masers are uncommon during star formation,
polarization observations are understandably rare, however,
observations of the 86~GHz SiO masers near the massive protostar Orion
IRc2 reveal a high level of fractional polarization and a magnetic
field along the disk (\cite[Plambeck et al. 2003]{Plambeck03}).

\subsection{OH masers}

\begin{figure}
 \centerline{
 \scalebox{0.35}{\includegraphics{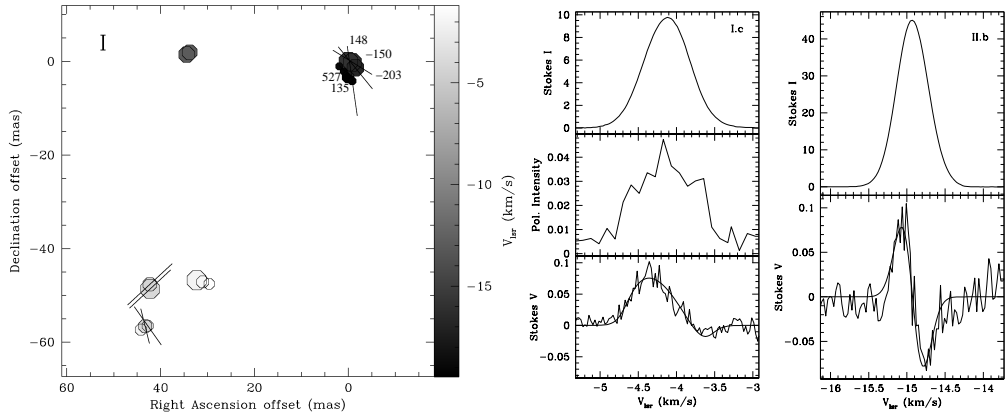}}}
  \caption{From \cite{Vlemmings06b}: (left) The polarization of the
    H$_2$O masers that are argued to be excited in a expanding
    shockwave through a rotated protostellar disk in Cepheus A~HW2
    (\cite[Gallimore et al. 2003]{Gallimore03}). The linear
    polarization vectors, scaled logarithmically according to
    polarization fraction $P_l$, are over-plotted. For the maser
    features where the Zeeman splitting was detected the magnetic
    field strength is indicated in mG. (right) Total power (I) and
    V-spectra for selected maser features in Cepheus A. Additionally,
    the linear polarized flux density, $\sqrt{(Q^2+U^2)}$, is shown when
    detected. The flux densities are given in Jy~beam$^{-1}$. The
    thick solid line in the bottom panel shows the best non-LTE model
    fit to the circular polarization V. The V-spectrum is adjusted by
    removing a scaled down version of the total power spectrum and are
    clearly not necessarily symmetric.}
\label{Fig:cepa}
\end{figure}

There exists a large number of OH maser polarization measurements,
probing densities from $\sim 10^5 - 10^8$~cm$^{-3}$. OH masers are
often strongly polarized and as the Zeeman splitting is large,
observations of the separate $\sigma^+$ and $\sigma^-$ components
directly yields a magnetic field strength. Observations of the $100\%$
linearly polarized $\pi$-component are extremely rare however, likely
due to magnetic beaming and the overlap of several differently
polarized masers along the line of sight (\cite[Fish \& Reid
  2006]{Fish06}). The measured field strengths are typically around
$\sim 1$~mG although recent observation have also revealed individual
maser spots with field strengths up to $\sim 40$~mG (\cite[Fish \&
  Reid 2007]{Fish07}). The field direction observed from OH masers
seems to maintain the direction of the ambient $B$-field as the
observations show a uniform direction of the field with at most one
reversal across regions as large as several arcmin
(e.g. \cite[Bartkiewicz et al. 2005]{Bartkiewicz05}). While linear
polarization measurements can sometimes be used to constrain the
three-dimensional $B$-field configuration
(e.g. \cite[Hutawarakorn et al. 2002]{Hutawarakorn02}, \cite[Gray et
  al. 2003]{Gray03b}), internal and external Faraday rotation make a
direct interpretation of the 1.6~GHz OH maser linear polarization
difficult.

Magnetic field measurements have also been made using several of the
excited OH maser transitions at 6 and 13~GHz. These observations suffer
less from Faraday rotation while tracing mostly the same regions of
the star forming region. The polarization and $B$-field values are
fully consistent with the lower excitation measurements
(e.g. \cite[Baudry \& Diamond 98]{Baudry98}, \cite[Caswell
  2003]{Caswell03}, \cite[Desmurs et al. 1998]{Desmur98}, \cite[Etoka
  et al. 2005]{Etoka05}).

\subsection{H$_2$O masers}

After the first discovery of interstellar H$_2$O maser Zeeman
splitting by \cite{Fiebig89} using single dish observations, there
have been an increasing number of higher spatial resolution circular
polarization observations confirming the earlier results
(e.g. \cite[Sarma et al. 2001, 2002]{Sarma01}). These observations
typically reveal $B$-field strengths between $15$ and $150$~mG at
densities of $n_{\rm H_2}=10^8 - 10^{11}$~cm$^{-3}$. The largest field
was recently found in a proposed protostellar disk around one of the
several protostars in the Cepheus~A~HW2 region, with a field strength
of $\sim 650$~mG, as shown in Fig.~\ref{Fig:cepa} (\cite[Vlemmings et
  al. 2006b]{Vlemmings06b}). Such high $B$-field strength implies a
nearby source enhancing the magnetic field. Considering the size of
the disk this implies a $B$-field strength of $\sim 2.5$~G near the
embedded protostar. Alternatively, the observations indicate much
higher densities then current H$_2$O maser theory allows in the
shockwave where the masers are excited (\cite[Elitzur et
  al. 1989]{Elitzur89}, and references therein).

In addition to the circular polarization, low levels of linear
polarization (typically $\lesssim 2\%$) are also observed in
star forming regions. While often structure in the $B$-field
direction is detected, the observations show rapid changes of
direction over small scales (see Fig.~\ref{Fig:cepa}).

\subsection{Methanol masers}

\begin{figure}
 \centerline{
 \scalebox{0.3}{\includegraphics{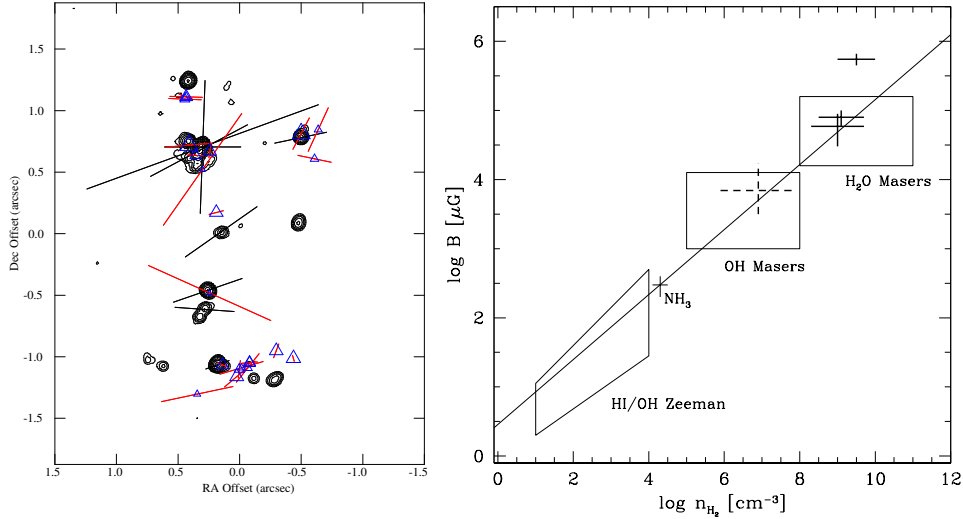}}}
  \caption{(left) The methanol masers of W3(OH) (contours) including
    the polarization vectors (black) scaled linearly according the
    fractional linear polarization (\cite[Vlemmings et
      al. 2006c]{Vlemmings06c}). The positions in the map are
    indicated with respect to R.A.(J200)$=02^h27^m03.7743^s$,
    Dec(J2000)$=61^\circ52'24.549''$. The (blue) triangles denote the
    main line OH masers from \cite{Wright04b} for which polarized
    intensity was detected at $5\sigma$ significance (polarized flux
    $>75$~mJy), and the grey (red) vectors are their linearly scaled
    polarization vectors. The main line OH maser polarization vectors
    lengths are scaled down by a factor of 5 with respect to the
    lengths of the methanol maser polarization vectors. (right) The
    magnetic field strength $B$ in massive star forming regions
    measured from Zeeman measurements as a function of $n_{\rm H_2}$,
    the number density of neutral hydrogen. The boxes indicate the
    literature values for HI/non-masing OH, OH maser and H$_2$O maser
    Zeeman splitting observations. Also indicated are the H$_2$O maser
    measurements (thick solid crosses) and OH maser measurements
    (thick dashed cross) in the star forming region Cepheus A
    (\cite[Vlemmings et al. 2006b]{Vlemmings06b}, \cite[Bartkiewicz et
      al. 2005]{Bart05}) as well as an NH$_3$ measurement for the same
    source from \cite{Garay96}. The solid line is the empirical
    relation $B\propto n^{0.49}$ determined by \cite{Crutcher99} fixed
    to the NH$_3$ magnetic field measurement.}
\label{Fig:fig4}
\end{figure}

Although the 6.7 and 12~GHz methanol masers are some of the most
abundant masers in high-mass star forming regions, polarization observations
have been rare. \cite{Ellingsen02} presented ATCA linear polarization
measurements of the 6.7~GHz masers, with a typical polarization
fraction $m_l \sim 1.5\%$, while \cite{Koo88} presented similar
polarization fractions for the 12.2~GHz maser. Finally,
\cite{Wiesemeyer04} claim up to $40\%$ linear polarization of the
millimeter methanol masers. Only recently has the first 6.7~GHz
methanol maser polarization map been made of the star forming region
W3(OH), as shown in Fig.~\ref{Fig:fig4}(left) (\cite[Vlemmings et
  al. 2006c]{Vlemmings0c}). These observations show that the $B$-field
is aligned with the large scale methanol maser filament and are
consistent with previous OH maser observations. Importantly, the
observations show that because they are less influenced by both
internal and external Faraday rotation, methanol masers are better
probes of the overall $B$-field structure than OH masers. Since the
Zeeman splitting of methanol is small and circular polarization
measurements are often hindered by dynamic range problems, the
observations of \cite{Vlemmings06c} only provide an upper limit to the
$B$-field strength. However, the first tentative detection of a
$B$-field strength from methanol observations has recently been made
(Green et al. 2007, this proceedings).

\subsection{Summary}


Fig.~\ref{Fig:fig4}(right) shows the $B$-field measurements of both
masers and non-maser observations as a function of number density. The
figure seems to indicate that (with the exception of regions such as
the above mentioned protostellar disk), the $B$-field follows the
density scaling law over an enormous range of densities, implying that
the magnetic field remains partly coupled to the gas up to the highest
number density (see Crutcher 2007, this proceedings). However, the
shock excited H$_2$O maser are short-lived (with a typical lifetime
$\tau_m \sim 10^8$~s) compared to the typical adiabatic diffusion
timescale at the highest densities ($\tau_d \sim 10^9$~s), implying
that in the non-masing gas of similar densities, magnetic field
strengths are likely lower due to the adiabatic diffusion. Still, the
maser $B$ measurements strongly imply a dynamical importance of
magnetic fields during the high-mass star formation process,
especially in shaping outflows and jets.

\section{Other sites of maser emission}
\label{sect4}

Besides the evolved stars and star forming regions, masers are found in
several other types of sources. Of those, both supernova remnant (SNR)
masers and megamasers have had polarization measurements reported in
the last few years.

\subsection{Supernova remnants}

The 1720~MHz OH masers in SNRs have been the target of several
polarization observations and recently high resolution observations
have for the first time directly resolved the Zeeman splitting between
right- and left-circularly polarized spectra (\cite[Hoffman et
  al. 2005, and references therein]{Hoffman05}). Linear polarization
reveals the post-SNR shock $B$-field is well ordered and
consistent with other magnetic field tracers and typical derived
$B$-field strengths are $|B| \sim 0.5-1.0$~mG (see Brogan 2007,
this proceedings).

\subsection{Extragalactic megamasers}

Finally, both H$_2$O and OH megamasers have been targeted for
polarization studies. As the accretion disk H$_2$O masers are weak,
only upper limits have been determined for the magnetic field strength
in the masing region. A limit of $30$~mG was found in the disk of
NGC~4258, which has been used to determine a black hole accretion rate
$< 10^{-3.7}$~M$_\cdot$/yr (\cite[Modjaz et al. 2005]{Modjaz05}). The
tightest limit found to date is $\sim 11$~mG in the disk of NGC~3079
at 0.64~pc from the central black hole (\cite[Vlemmings et
  al. 2007]{Vlemmings07}), while a further limit of $\sim 150$~mG was
found for the $B$-field in the Circinus H$_2$O maser disk
(\cite[McCallum et al. 2007]{McCallum07}).

The first actual measurement of a surprisingly strong extragalactic
maser magnetic field strengths was recently made by Robishaw et
al. (2007, this proceedings) on the 1720~MHz OH masers in a number of
ULIRG.

\section{Future perspectives}
\label{sect5}

Since the previous IAU maser conference (Migenes \& Reid 2002) there
has been a huge increase in maser polarization measurements. H$_2$O
maser polarization observations have become much more common, the
first extragalactic maser polarization detections have been made and
methanol maser polarization observations have shown their
enormous potential. With the advent of new instruments such as the
SMA and ALMA, high resolution dust polarization observations will soon
be able to bridge the gap between the small scale maser magnetic field
measurements and the very large scale single dish dust polarization
observations. This should finally be able to convincingly show the
importance of maser polarization observations in star forming regions
for magnetic field studies, where especially methanol masers will be
able to play a hugely important role. Further in the future, the SKA
should be able to bring H$_2$O maser observations to a new level,
fully imaging the polarization of the H$_2$O envelopes of evolved
stars, PNe and even AGN accretion disks. But already in the near
future, existing and soon to be upgraded instruments such as (e)MERLIN,
the (e)VLA and the VLBA will be able to answer the questions about the
role of magnetic fields during star formation and late stellar
evolution by further increasing the size of the source samples.


\end{document}